\documentclass[final,authoryear,5p,times,twocolumn]{elsarticle}

\usepackage{amssymb}
\usepackage{amsmath}
\usepackage{amssymb}
\usepackage{graphicx}
\usepackage{longtable}

\biboptions{authoryear,semicolon}

\journal{Biosystems}

\begin{document}

\begin{frontmatter}

\title{Positive feedback and temperature mediated molecular switch controls differential
gene regulation in {\it Bordetella pertussis}
}

\author{Arnab Bandyopadhyay}

\author{Suman K Banik\corref{cor}}
\ead{skbanik@bic.boseinst.ernet.in}

\cortext[cor]{Corresponding author; Phone: +91-33-2303 1142; Fax: +91-33-2303 6790}

\address{Department of Chemistry, Bose Institute, 93/1 A P C Road, Kolkata 700 009, India}%

\begin{abstract}
Based on the phosphorelay kinetics operative within BvgAS two component system we
propose a mathematical framework for signal transduction and gene regulation of 
phenotypic phases in {\it Bordetella pertussis}. The proposed model identifies a novel 
mechanism of transcriptional interference between two promoters present in the {\it bvg} 
locus. To understand the system behavior under elevated temperature, the developed 
model has been studied in two different ways. First, a quasi-steady state analysis has been 
carried out for the two component system, comprising of sensor BvgS and response regulator 
BvgA. The quasi-steady state analysis reveals temperature induced sharp molecular switch, 
leading to amplification in the output of BvgA. Accumulation of a large pool of BvgA thus results 
into differential regulation of the downstream genes, including the gene encoding toxin. 
Numerical integration of the full network kinetics is then carried out to explore time dependent behavior of different system components, that qualitatively capture the essential features of  experimental results performed {\it in vivo}. Furthermore, the developed model has been utilized 
to study mutants that are impaired in their ability to phosphorylate the transcription factor, 
BvgA, of the signaling network.
\end{abstract}


\end{frontmatter}


\section{Introduction}

One of the important functional aspects of living organisms is to respond to the sudden
changes made in their environment, and to make appropriate changes in the cellular or
subcellular level for survival.
Direct manifestations of such changes at the subcellular level
are the expression/repression of single or multiple genes controlling different functional
behavior of an organism \citep{Alon2007}.
To achieve this, living system utilizes concerted biochemical network composed of 
several feedback mechanism \citep{Tyson2003,Tyson2010}.
The human pathogen {\it Bordetella pertussis}, a gram
negative bacteria and causative agent for the disease whooping cough \citep{Preston2004},
is no exception to the aforesaid behavior. At 25 $^{\circ}$C, while freely moving in the
environment their
pathogenic properties remain dormant. But, when they are within the host at 37 $^{\circ}$C,
their virulent properties come into play. In the laboratory the reverse effect, i.e., suppression
of pathogenic behavior is observed using MgSO$_4$ or nicotinic acid
\citep{Beier2008,Cotter2003}. The virulent
behavior of {\it B. pertussis} within host, in response to sudden environmental change,
has been experimentally studied and has been found to be operative through BvgAS
two component system (TCS) \citep{Beier2008,Cotter2003}.  The TCS
comprises of transmembrane sensor BvgS and response regulator BvgA where signal flows
through this pair via a four step (His-Asp-His-Asp) phosphorelay mechanism.

As a response to temperature elevation in the environment, the response regulator
BvgA  becomes active (the phosphorylated dimer) within each bacterium, which in turn
exerts a positive feedback on its own operon, the {\it bvg} operon.
Positive feedback loop thus increases the active form of BvgA in a switch like manner. 
In other words, once the BvgAS two-component machinery becomes operative, large pool 
of active BvgA either repress and/or express several downstream genes where the 
phosphorylated dimer of BvgA plays the leading role by acting as transcription factor (TF) 
\citep{Steffen1996}. In the laboratory condition at 37 $^{\circ}$C and in the absence of
MgSO$_4$ or nicotinic acid BvgA activates transcription of virulence activated genes
(\textit{vag}), as well as represses transcription of virulence repressed genes (\textit{vrg}).
Due to this reason \textit{bvg} locus was earlier called \textit{vir} due to its connection
to virulence \citep{Beier2008,Cotter2003,Weiss1984}.
In \textit{B. pertussis}, \textit{vrg} loci encodes outer membrane whereas in
\textit{B. bronchiseptica}, \textit{vrg} controlled genes encode motility and survival from 
nutrient limitation condition. In \textit{Bordetella spp.}, \textit{vag} loci encodes genes 
responsible for adherence, toxins (including pertussis toxin in \textit{B. pertussis}, a type 
III secretion system and BvgAS itself (due to autoregulation).

Based on the binding affinity of TF to the respective promoters, different types
of downstream genes are regulated in {\it Bordetella spp.} and have been broadly grouped
into four classes, e.g., class 1, class 2, class 3 and class 4 \citep{Beier2008,Cotter2003}.
Class 1 genes encompass genes that are responsible for encoding toxins,
such as adenylate cyclase ({\it cyaA-E}) and pertussis toxin ({\it ptxA-E}). Class 2
genes express proteins responsible for adherence, such as {\it fhaB} that encodes
filamentous hemagglutinin. Among all the four classes of genes, class 3 genes show a unique
behavior,  although its functional activity is not known till date \citep{Beier2008,Cotter2003}.
The only well characterized class 3 gene found in {\it B. pertussis}
is known as {\it bipA}. The final one, class 4 genes have been reported to encode
{\it frlAB} in {\it B. bronchiseptica} and is responsible for motility.
It is important to mention that expressions of class 3 and class 4 gene are not observed in
{\it B. pertussis} under the influence of temperature elevation. To be specific, class 3
gene expression has been observed in {\it B. pertussis}  only under the influence
of intermediate concentration of MgSO$_4$ and class 4 gene under low concentration of
MgSO$_4$ in {\it B. bronchiseptica} \citep{Beier2008,Cotter2003}.
In terms of \textit{vir} regulated genes class 4 gene thus belongs to \textit{vrg}
whereas class 1 and class 2 genes belong to \textit{vag}.
Expression and/or repression of the four classes of downstream genes is controlled by
strong and/or weak binding sites (for TF) present in the promoter region of the respective
genes. Among these, promoter region of class 4 gene has the strongest affinity for TF.
Promoter region of class 2 and class 3 genes have medium affinity for TF, whereas promoter
region of class 1 gene has the weakest affinity for TF. On the basis of the promoter
regions' affinity for TF it is thus expected that expression and/or repression of four classes
of downstream genes in {\it Bordetella spp.} would show a differential pattern in their
temporal dynamics.

Keeping these aforesaid phenomenological information in mind we have developed a
mathematical model based on biochemical interactions taking place within {\it B. pertussis}
under the influence of temperature elevation. The objective of present work is twofold.
First, we aim to understand the molecular switch operative in BvgAS TCS and to identify
the key players responsible for amplification of TF. Second, through our model we aim to
regenerate qualitative features of the network and to mimic different phenotypic
states of {\it B. pertussis} under temperature elevation, as well as their expression level
due to different mutation.


\begin{figure}[!t]
\begin{center}
\includegraphics[width=1.0\columnwidth,angle=0]{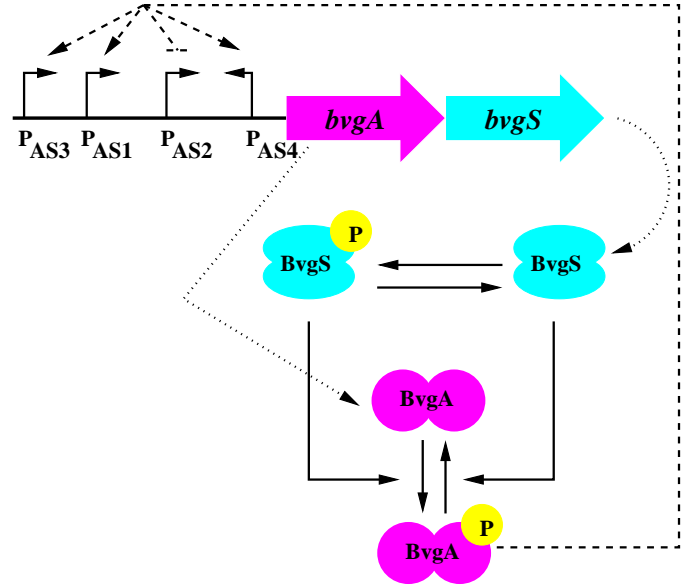}
\end{center}
\caption{(color online) Schematic presentation of {\it bvg} locus and signal transduction in BvgAS
two component system. The dashed line presents the feedback by phosphorylated dimer
of BvgA on its own operon. The dotted line is for the production of dimers
of BvgS and BvgA. For simplicity the mRNAs are not shown in the diagram.
}
\label{cartoon1}
\end{figure}


\section{The Model}

To understand the mechanism for temperature induced activation of {\it bvg} locus
and differential regulation of the downstream genes, we propose a kinetic model in 
the following.


\subsection{The {\it bvg} locus}

Experimental studies in {\it B. pertussis} suggest multi-promoter activities in
{\it bvg} operon \citep{Roy1990,Scarlato1990,Scarlato1991}.
Out of the four promoters, $P_{AS1}$, $P_{AS2}$, $P_{AS3}$ and $P_{AS4}$, 
present in the {\it bvg} locus (see Figure~\ref{cartoon1}), only $P_{AS2}$ is known to 
be constitutively active under non-inducing condition (25 $^{\circ}$C) and is {\it bvg} 
independent. After induction (37 $^{\circ}$C), activity of the $P_{AS2}$ promoter goes 
down while the other three promoters ($P_{AS1}$, $P_{AS3}$ and $P_{AS4}$)  become 
active. As shown in \cite{Scarlato1991}, at 37 $^{\circ}$C, $P_{AS1}$ shows maximal level 
of activity compared to $P_{AS3}$ and is on within $< 10$ minutes of induction. The amount 
of transcripts generated from $P_{AS3}$ is very low and have been reported to be hardly 
detectable. The $P_{AS4}$ promoter shows same level of activity as $P_{AS1}$ but produces antisense RNA. Although activity of $P_{AS4}$ promoter and its product, the antisense-RNA, 
is known, the target of the antisense RNA is not known till date. In passing it is important to 
mention that multi-promoter activity in the operon of TCS as observed in {\it B. pertussis}, has also
been observed in other human pathogens \citep{Chauhan2008,Dona2008}.

To study functioning of the {\it bvg} locus we consider only the activity of two the
promoters $P_{AS1}$ and $P_{AS2}$ in our model, as reasonable amount of experimental
data is available in the literature for these two promoters \citep{Scarlato1991}.
Both these promoters are typical example of tandem promoter,
containing a conserved region of $\approx 10$ base pairs between upstream of
$P_{AS1}$ and transcriptional start site (TSP) of $P_{AS2}$.  
In the model, we designate the constitutive
form of $P_{AS2}$ under non-inducing condition as $P_{AS2c}$. Once induced,
TF interacts with both $P_{AS1}$ and $P_{AS2}$ and makes them active
\begin{eqnarray}
\label{locus}
P_{AS2c} + A_{2P}
& \overset{k_{b2}}{\underset{k_{u2}}{\rightleftharpoons}} &
P_{AS2a} , \\
P_{AS1i} + A_{2P}
& \overset{k_{b1}}{\underset{k_{u1}}{\rightleftharpoons}} &
P_{AS1a} .
\end{eqnarray}


\begin{figure}[!t]
\begin{center}
\includegraphics[width=1.0\columnwidth,angle=0]{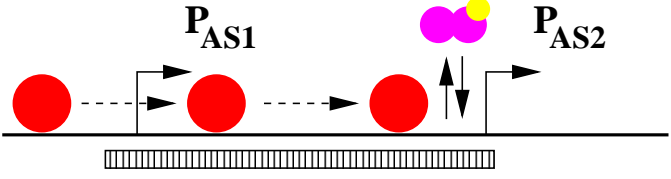}
\end{center}
\caption{(color online) Schematic presentation of transcriptional interference between 
$P_{AS1}$ and $P_{AS2}$.  RNA polymerase (red blob) starts its journey from upstream 
of $P_{AS1}$
promoter executing initiation, elongation and termination (follow the dotted arrowhead).
During termination it interferes with the TF ($A_{2P}$, magenta dimers with yellow blob
on top), causing transcriptional interference and downregulation of $P_{AS2}$ promoter
activity. The vertical solid arrowheads presents binding/unbinding process between TF
and $P_{AS2}$ promoter site. The filled bar at the bottom is for overlapping $\approx 10$
base pair region between upstream of $P_{AS1}$ and TSP of $P_{AS2}$.
}
\label{cartoon2}
\end{figure}

\noindent
In the above equations $P_{AS2a}$ is the active form of $P_{AS2}$;
and $P_{AS1i}$ and $P_{AS1a}$ are inactive and active form of $P_{AS1}$ promoter, 
respectively.
Due to the presence of conserved region of $\approx 10$ base pairs, RNA polymerase 
(RNAP) for $P_{AS1}$ traversing through downstream of $P_{AS1}$ now interferes with 
the binding 
of TF (and RNAP for $P_{AS2}$) to upstream of $P_{AS2}$ causing transcriptional interference
(see Figure~\ref{cartoon2})
\citep{BuettiDinh2009,Shearwin2005}. During this process 
$P_{AS2}$ and $P_{AS1}$ act as 
sensitive and aggressive promoter, respectively. Although a detailed kinetic mechanism 
of  transcriptional interference has been proposed and verified experimentally 
\citep{BuettiDinh2009}, we use the following notion to keep the model simple
\begin{equation}
\label{p2ti}
P_{AS2a}
\overset{k_{i2}}{\underset{k_{a2}}{\rightleftharpoons}}
P_{AS2i} ,
\end{equation}

\noindent
where the information of upstream inhibition are put together in the rate constants.
The rate constant $k_{i2}$ in Eq.~(\ref{p2ti}) contains the information of RNAP coming 
from $P_{AS1}$ causing  transcriptional interference at $P_{AS2}$,
\begin{eqnarray*}
P_{AS2a} +RNAP
\overset{k_{i2}^{\prime}}{\underset{k_{a2}}{\rightleftharpoons}}
P_{AS2i} .
\end{eqnarray*}

\noindent
Considering the pool of RNAP to be constant one can absorb it into the overall rate
of the interference mechanism and write $RNAP \; k_{i2}^{\prime} = k_{i2}$, the overall
rate constant of the interference process given in Eq.~(\ref{p2ti}). After causing the
interference the RNAP coming from $P_{AS1}$ falls off from $P_{AS2}$ thus giving
chance to the later to back to the active form again, which is modeled as the backward
reaction with rate constant $k_{a2}$ in Eq.~(\ref{p2ti}). This helps the $P_{AS2}$
promoter to maintain a low activity even after 2 hours of induction as evident from the
experimental data (see Figure 3B of \cite{Scarlato1991} and Figure~\ref{mrna} of the present
work).
To keep track of the transcripts generated due to $P_{AS1}$ and $P_{AS2}$, following
\cite{Scarlato1991}, we consider two isoforms of mRNA generated from the
{\it bvg} locus, $m_{AS1}$ and $m_{AS2}$, respectively. At 25 $^{\circ}$C, $m_{AS2}$
is constitutively produced from the constitutive state of $P_{AS2}$ promoter
\begin{subequations}
\begin{equation}
\label{m21}
P_{AS2c} \overset{k_{tp,20}}{\longrightarrow} m_{AS2} .
\end{equation}

\noindent
After 2 hours of induction, level of $m_{AS2}$ goes down but still maintains a low level
of expression (see Figure 3B of \cite{Scarlato1991}) which we assign to the
residual activity of $P_{AS2a}$. After 20 minutes of induction, production of $m_{AS2}$ 
remains still on, reaches a maxima, and then goes down which we attribute to the 
suppression of $P_{AS2a}$ and the formation of $P_{AS2i}$ altogether. Thus after induction,
\begin{equation}
\label{m22}
P_{AS2a} \overset{k_{tp,21}}{\longrightarrow} m_{AS2} ,
\end{equation}
\end{subequations}

\noindent
Unlike the transcripts generated due to $P_{AS2}$ activity, generation of $m_{AS1}$ is solely
governed by the active form of $P_{AS1}$ promoter
\begin{equation}
\label{m11}
P_{AS1a} \overset{k_{tp,11}}{\longrightarrow} m_{AS1} .
\end{equation}

\noindent
Finally, we consider natural degradation of both the transcripts generated from 
$P_{AS1}$ and $P_{AS2}$,
\begin{equation}
m_{AS1} \overset{k_{d,m}}{\longrightarrow} \varnothing ,
m_{AS2} \overset{k_{d,m}}{\longrightarrow} \varnothing .
\end{equation}


\subsection{The two component system}

Once transcribed from the locus, we consider translation of two isoforms $m_{AS1}$
and $m_{AS2}$ into dimers of sensor, BvgS ($S_2$), and response regulator,
BvgA ($A_2$) proteins,
\begin{subequations}
\begin{eqnarray}
\label{synth-s}
m_{AS1} \overset{k_{ss,1}}{\longrightarrow} m_{AS1} + S_2 , \\
m_{AS2} \overset{k_{ss,2}}{\longrightarrow} m_{AS2} + S_2 , \\
\label{synth-a}
m_{AS1} \overset{k_{sa,1}}{\longrightarrow} m_{AS1} + A_2 , \\
m_{AS2} \overset{k_{sa,2}}{\longrightarrow} m_{AS2} + A_2 .
\end{eqnarray}
\end{subequations}

\noindent
In reality the sensor and response regulator proteins are first translated as monomers
and then dimerize \citep{Beier2008}. Since dimers of BvgA, not the monomers,
act as transcription factors for the activation of {\it bvg} locus, we have omitted kinetics
of monomer formation and subsequent dimerization in our model and work instead with
$S_2$ and $A_2$. As we show in the following, this simplification does not affect our analysis
and modeling of the signal transduction network.

In bacterial TCS, autophosphorylation occurs at histidine residue of the
sensor kinase, that serves as source of phosphate group and transfers the same to the
aspartate residue of response regulator, acting as sink
\citep{Appleby1996,Hoch2000,Laub2007,Mitrophanov2008,Stock2000}.
This orthodox two-step His-Asp phosphotransfer becomes complicated in {\it B. pertussis}
where signal transduction takes place via a unorthodox four-step His-Asp-His-Asp
phosphotransfer mechanism \citep{Uhl1996},
where the first three steps (His-Asp-His) take place within the sensor protein BvgS and in
the last step (His-Asp) phosphate group flows from the transmembrane sensor BvgS
to the cytoplasmic response regulator BvgA. 
Mathematical modeling of two-step phosphorelay has been
reported in different context of bacterial signal transduction
\citep{Banik2009,Batchelor2003,Kato2007,Kierzek2010,Kremling2004,Shinar2007,Sureka2008}.
Whereas mathematical modeling of four-step phosphorelay has been extensively used
for the gram positive bacteria \textit{Bacillus subtilis} to understand the sporulation initiation,
a detailed account of which can be found in the recent review by \cite{Liebal2010}.
In \textit{B. subtilis} the external signal is sensed by a family of five histidine kinase 
KinA-E and finally transferred to the response regulator Spo0A via Spo0F and Spo0B. Phosphorylated Spo0A acts as a TF by controlling over 500 genes \citep{Fawcett2000} 
which are broadly classified into two categories by the affinity of Spo0A to their target genes \citep{Fujita2005}. In passing we would like to mention the work by \cite{Kim2006} where
a comparative study of two-step versus four-step phosphorelay has been undertaken.
In the present study we have adopted a simplified approach of what is proposed by 
\cite{Kim2006}. Instead of detailed three-step (His-Asp-His) phosphotransfer mechanism 
within BvgS we consider autophosphorylation at $S_2$, the dimer of BvgS,
\begin{equation}
\label{autop}
S_2 \overset{k_{p,s2}(s)}{\underset{k_{dp,s2}}{\rightleftharpoons}} S_{2P}.
\end{equation}

\noindent
While writing the above equation we have put the information of ATP
and its interaction with the sensor protein in the rate constant $k_{p,s2} (s)$;
In addition it is a function of input signal $s$ which may be temperature or 
salt concentration for {\it B. pertussis}. In absence of any signal, $s=0$,
autophosphorylation at the histidine residue becomes nonfunctional,
i.e., $k_{p,s2} (0)=0$.
At this point it is important to mention that exact mechanism for the activation of
BvgS in {\it B. pertussis} under temperature induction is not clear from the
literature. To incorporate sensing of the external stimulus and subsequent
activation of the TCS we have adopted the mechanism given in \ref{autop}.
Once phosphorylated the sensor protein transfers the phosphate group to their
cognate response regulator $A_2$, the dimer of BvgA, mimicking the last step (His-Asp)
of the four-step phosphotransfer mechanism
\begin{equation}
\label{ptrans}
S_{2P} + A_2 \overset{k_{t,f}}{\underset{k_{t,b}}{\rightleftharpoons}}
S_{2P}\cdot A_2 \overset{k_{t,a2}}{\longrightarrow}
S_2 + A_{2P} ,
\end{equation}

\noindent
where $S_{2P}\cdot A_2$ is the Michaelis complex formed by $S_{2P}$ and $A_2$.
In addition to their kinase activity the sensor protein also exhibits phosphatase activity 
by removing the phosphate group from their cognate partner \citep{Batchelor2003,Laub2007,Shinar2007}
thus showing bifunctional behavior
\begin{equation}
\label{pphos}
S_2 + A_{2P} \overset{k_{p,f}}{\underset{k_{p,b}}{\rightleftharpoons}}
S_2\cdot A_{2P} \overset{k_{p,a2}}{\longrightarrow}
S_2 + A_2 .
\end{equation}

\noindent
Likewise the mRNA transcribed from the {\it bvg} locus, we consider natural degradation
of different forms of sensor and response regulator
proteins
\begin{eqnarray}
S_2 \overset{k_{d,p}}{\longrightarrow} \varnothing,
S_{2P} \overset{k_{d,p}}{\longrightarrow} \varnothing,
A_2 \overset{k_{d,p}}{\longrightarrow} \varnothing,
A_{2P} \overset{k_{d,p}}{\longrightarrow} \varnothing .
\end{eqnarray}

\noindent
The phosphorylated dimer of the response regulator then exerts a positive feedback in the
{\it bvg} locus thus activating the promoters of the locus (see Eqs.~(1-2)).
In addition they control the expression and/or repression of several downstream genes.


\begin{figure}[!t]
\begin{center}
\includegraphics[width=1.0\columnwidth,angle=0]{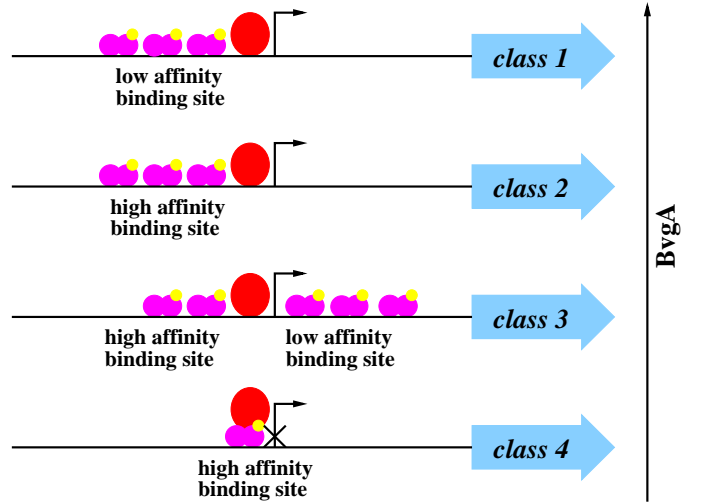}
\end{center}
\caption{(color online) Phenotypic gene regulation in
{\it B. pertussis}. The red blob is for RNA polymerase and the magenta
dimers with yellow blob on top are for phosphorylated dimers of BvgA.
}
\label{cartoon3}
\end{figure}


\subsection{Differential gene regulation}

To model regulation of downstream genes under induced condition we use the following
TF binding properties of different promoters of downstream genes.
The promoters for class 1 gene contain low affinity BvgA binding site
far upstream of the TSP, as a result high level
of $A_{2P}$ is necessary to activate these genes and they are expressed quite
late compared to class 2 and class 3 genes.
Class 2 promoters contain high affinity
BvgA binding site close to TSP and fairly low level of $A_{2P}$ is sufficient to
activate these genes. As a result, expression of class 2 genes becomes visible
within very short period after induction.
Class 3 gene, {\it bipA}, contains
high affinity as well as low affinity BvgA binding site just upstream and
downstream of TSP, respectively. Once induced, {\it bipA} becomes active almost
at the same time like {\it fhaB} (class 2 gene) with the help of low amount of
$A_{2P}$. At a critical concentration of BvgA gene expression becomes maximum
and then it starts falling due to BvgA binding to the downstream low affinity
binding site.
The {\it frlAB} promoter (class 4) has been found to contain distinguishable BvgA
binding site that overlaps
with TSP \citep{Akerley1995}. Thus a very low level of BvgA may be able to suppress
class 4 genes.

In Figure~\ref{cartoon3} we schematically show regulation of four classes of genes as
BvgA level increases. The binding kinetics of phosphorylated dimer of BvgA
($A_{2P}$) to the promoters of these four classes of genes are given in the
Supplementary data. While active, four classes of genes starts
transcribing their corresponding mRNA
\begin{equation}
\label{mcl1}
P_{clj,a} \overset{k_{tp,clj}}{\longrightarrow} m_{clj} ,
\end{equation}

\noindent
where $j=1-4$. Likewise the transcripts generated from the {\it bvg} locus,
we consider natural degradation of the four different classes of transcripts
\begin{eqnarray}
m_{clj} \overset{k_{d,m}}{\longrightarrow} \varnothing .
\end{eqnarray}

\noindent
Before proceeding further we would like to mention that all the relevant symbols
designating biochemical species used in this section are listed in Table~1.


\section{Results and Discussions}

To check the validity of our proposed model, the kinetic network developed in the previous
section has been translated into sets of coupled nonlinear ordinary differential equations
(ODEs). To understand functionality
of temperature induced switch operative in the proposed model we first analyze the kinetics for
{\it bvg} locus and the TCS using quasi-steady state approximation \citep{Batchelor2003}.
The full network kinetics is then numerically integrated for large set of parameters of the
proposed model and compared with {\it in vivo} experimental results \citep{Scarlato1991}.


\subsection{Steady state analysis}

According to the model described in the previous section total amount of sensor and
response regulator proteins can be expressed by the following relations
\begin{eqnarray}
\label{stot}
[S_T] & = & [S_2] + [S_{2P}] + [S_{2P}\cdot A_2] + [S_2\cdot A_{2P}] , \\
\label{atot}
[A_T] & = & [A_2] + [A_{2P}] + [S_{2P}\cdot A_2] + [S_2\cdot A_{2P}] ,
\end{eqnarray}

\noindent
where $[A_T]/[S_T] \approx k_{sa,1}/k_{ss,1} \approx 6$ (see Table~2).
Thus one can express essential features of the dynamics in terms of the response regulator
protein. To realize the nature of amplification in gene expression at 37 $^{\circ}$C and to
analyze the steady state dynamics we divide the TCS signaling network into two modules, the
autoregulation module and the phosphorylation (or post-translational) module.

Before proceeding further we would like to comment on the input-output relation in the 
present model. For this, we define the influx and outflux of phosphate group 
in the network as 
\begin{eqnarray*}
J_i = k_{p,s2}(s) [S_2] \; \; {\rm and} \; \; J_o = k_{p,a2} [S_2 \cdot A_{2P}],
\end{eqnarray*}

\noindent
respectively \citep{Shinar2007}. Using the steady state expression of $[S_2 \cdot A_{2P}]$ (see
Eq.~(35) of the Appendix \textit{5.4}) and considering $J_i=J_o$ at steady state,
one arrives at $[A_{2P}^{ss}] = k_{p,s2}(s) K_{Mp}/k_{p,a2}$. Thus, the steady state
output of TF in the present model is dependent on the input signal and the phosphatase
activity of the sensor kinase on the response regulator. For further analysis we have removed 
the superscript $ss$ from the steady state expressions for notational simplicity.


\begin{figure}[!t]
\begin{center}
\includegraphics[width=0.75\columnwidth,angle=-90]{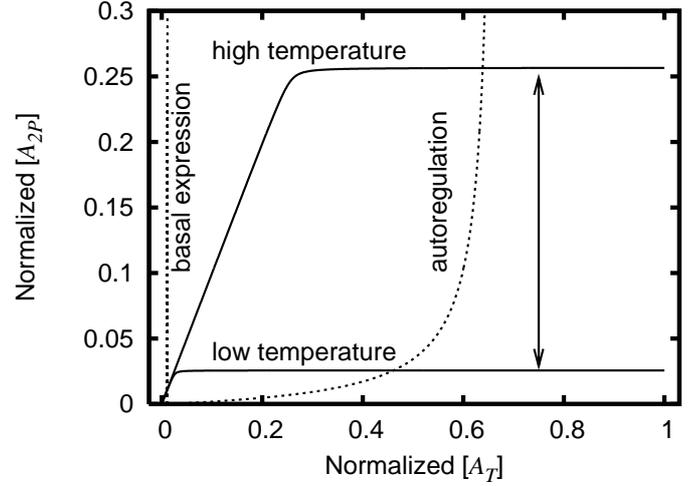}
\end{center}
\caption{The steady state behavior of the TCS  circuit in terms of normalized
$[A_{2P}]$ as a function of normalized $[A_T]$. Normalization was done with respect
to the total amount of response regulator at steady state. The solid curves are from 
phosphorylation 
module, Eq.~(\ref{php2}), at low and high temperature. The dotted switch like curve is 
due to autoregulation module, (Eq.~(\ref{aut2})). While going from low to high temperature,
a small change in the $[A_T]$ value makes a large amplification (the solid line with double 
headed arrow) in the $[A_{2P}]$ value (see the discussion in the main text). The leftmost 
vertical dotted line is for basal expression.
}
\label{phase}
\end{figure}


\subsubsection{Autoregulation module}

Autoregulation module takes care of synthesis and degradation of TF mediated by
positive feedback loop operative within \textit{bvg} operon. This ultimately leads into
total amount of response regulator, $A_T$, expressed in terms of the TF, $A_{2P}$,
in the steady state. To derive such expression we take time derivative of Eq.~(\ref{atot})
and impose quasi-steady state condition on $[S_{2P}\cdot A_2]$ and $[S_2\cdot A_{2P}]$
(see Appendix~\textit{5.4}) thus leading into
\begin{eqnarray}
\label{aut1}
0 & \approx & \tilde{\beta}_1 F_2 (A_{2P})
+ \tilde{\beta}_2 F_2 (A_{2P}) [A_{2P}]_2 \nonumber \\
&& +  \tilde{\beta}_3 F_1 (A_{2P}) - k_{d,p} [A_T] ,
\end{eqnarray}

\noindent
which finally yields the desired functional relationship between $A_T$ and $A_{2P}$
\begin{eqnarray}
\label{aut2}
[A_T] & \approx & \beta_1 F_2 (A_{2P})
+ \beta_2 F_2 (A_{2P}) [A_{2P}]_2 \nonumber \\
&& +  \beta_3 F_1 (A_{2P}) ,
\end{eqnarray}

\noindent
where $\beta_i = \tilde{\beta}_i/k_{d,p}$ $(i=1,2,3)$. The first two terms on the right hand side of
Eq.~(\ref{aut2}) arise due to $P_{AS2}$ promoter activity whereas the third term is solely due to
$P_{AS1}$ promoter under inducing condition. In the limit of zero phosphorylation
($k_{p,s2}=0$), i.e., at 25 $^{\circ}$C, Eq.~(\ref{aut2}) leads to
$[A_T] = \beta_1$  $(= k_{sa,2} k_{tp,20}/k_{d,m} k_{d,p})$. Thus, under zero stimulus
system dynamics is solely governed by basal transcription ($k_{tp,20}$) and translation
($k_{sa,2}$) processes.


\subsubsection{Phosphorylation module}

Considering only the phosphotransfer kinetics (see Eqs.~(\ref{autop}-\ref{pphos}))
and using the relations given in the Eq.~(35) of Appendix~\textit{5.4} we have at 
steady state
\begin{eqnarray}
\label{blns}
k_{p,s2} [S_2] = k_{dp,s2} [S_{2P}] - \frac{k_{t,a2}}{K_{Mt}} [S_{2P}] [A_2] , \\
\label{blna}
\frac{k_{p,a2}}{K_{Mp}} [S_2] [A_{2P}] =\frac{k_{t,a2}}{K_{Mt}} [S_{2P}] [A_2] .
\end{eqnarray}

\noindent
While deriving the above two expressions we have again imposed quasi-steady
state condition on the two Michaelis intermediates $[S_{2P}\cdot A_2]$ and 
$[S_2\cdot A_{2P}]$. After some algebra Eqs.~(\ref{blns}-\ref{blna}) provide
\begin{equation}
\label{php1}
\frac{C_p}{A_{2P}} = \frac{C_t}{A_2} + 1 ,
\end{equation}

\noindent
where $C_p = k_{p,s2}(s) K_{Mp}/k_{p,a2}$ and $C_t = k_{dp,s2} K_{Mt}/k_{t,a2}$.
Now, for $[A_T] \approx  [A_2] + [A_{2P}]$, Eq.~(\ref{php1}) gets transformed into
\begin{eqnarray}
\label{php2}
[A_{2P}] & \approx & \frac{1}{2} \left ( C_t + C_p + [A_T] \right ) \nonumber \\
&& - \frac{1}{2} \sqrt{\left ( C_t + C_p + [A_T] \right )^2 - 4 C_p [A_T]} ,
\end{eqnarray}

\noindent
which is valid for $A_{2P} \leqslant A_T$. Thus steady state output of $A_{2P}$
depends on both $A_T$ and the source of autophosphorylation, $k_{p,s2}(s)$. 
For $A_T  \gg C_t+C_p$, Eq.~(\ref{php2}) yields 
$A_{2P} \approx C_p$. Since $C_p$ is a function of $k_{p,s2}(s)$, for fixed set of 
other parameters of the model, a change in $k_{p,s2}(s)$ brings in a change in the 
value of $A_{2P}$. In addition, $A_{2P}$ level does not depend on $A_T$ anymore 
but remains dependent on ATP level through the rate constant $k_{p,s2}(s)$.

In Figure~\ref{phase} we show functional relation between normalized $[A_{2P}]$
and normalized $[A_T]$ at steady state \citep{Miyashiro2008}. Normalization was done 
with respect to the total amount of response regulator (phosphorylated and
unphosphorylated) at steady state. At steady state, depending on the level of 
input stimulus (here temperature of the environment), total amount of BvgA protein, 
$A_T$, exists either in phosphorylated form ($A_{2P}$) or in un-phosphorylated 
form ($A_2$) of which only the phosphorylated form acts as TF. The level of $A_{2P}$ 
at steady state is controlled by both the external temperature and the total amount of 
available BvgA protein. At low temperature only a small amount of $A_T$ gets 
transformed into $A_{2P}$ which increases as the temperature of the environment 
is raised. Enhancement of $A_{2P}$ level out of the total $A_T$ pool due to temperature
elevation is shown by the two sigmoidal solid curves in Figure~\ref{phase} using
Eq.~(\ref{php2}). It is interesting to note that even when a large pool of $A_T$ is
available to be phosphorylated only a small amount gets transformed into $A_{2P}$
at low temperature due to low autophosphorylation rate ($k_{p,s2}$) of Bvgs. But, at 
high temperature high $k_{p,s2}$ value changes the scenario by increasing the 
$A_{2P}$ level (see the discussion on amplification in subsection \textit{3.2}).

Formation of pool of $A_{2P}$ due to temperature elevation does not work due
to phosphotransfer mechanism only. It works hand in hand with the autoregulation 
module as well. As mentioned earlier, in the absence of external stimulus 
($k_{p,s2}=0$) the only form of BvgA available is $A_2$ that leads to $A_T=\beta_1$.
This gives the basal expression level (the left most vertical dotted line) 
in Figure~4. As the system gets switched on ($k_{p,s2} \neq 0$) $A_{2P}$
gets generated due to phosphotransfer and provides a positive feedback on 
the \textit{bvg} operon. Positive feedback enhances the production of $A_2$
which are ready to be phosphorylated at a fixed stimulus. This essentially increases
the amount of $A_{2P}$ out of $A_T$. This functional relation between $A_{2P}$ and
$A_T$ due to autoregulation is given by Eq.~(\ref{aut2}) and is plotted in
Figure~4 for nonzero $k_{p,s2}$ (the dotted switch like curve). The intersection
of the dotted curve and the solid curve at a particular temperature (low or high) gives 
the maximum limit of available $A_T$ that are available to be phosphorylated. Although, 
the difference between the intersection at low and high temperature shows that
due to small change in $A_T$ there is a huge change
in available $A_{2P}$. Thus, the intersections of the dotted curve due to autoregulation 
module and the solid lines due to phosphorylation module give a qualitative idea of
amplification in the steady state output of $A_{2P}$ when temperature of the surrounding 
is increased.


\begin{figure}[!t]
\begin{center}
\includegraphics[width=0.75\columnwidth,angle=-90]{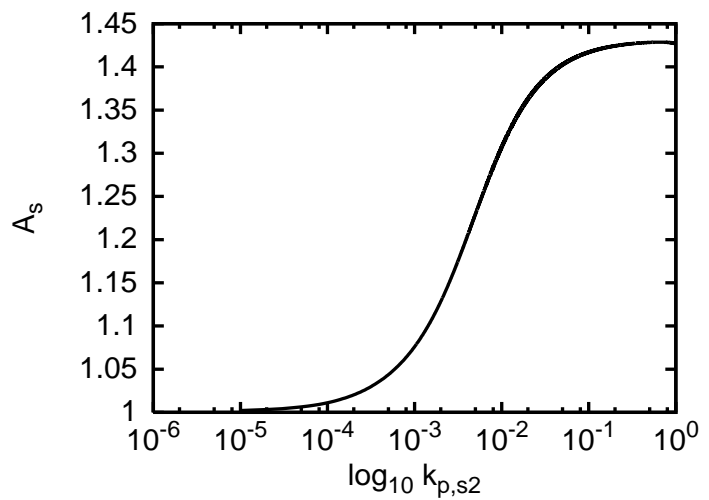}
\end{center}
\caption{Semilog plot of amplification factor, $A_s$ as a function of input 
stimulus, $k_{p.s2}$.
}
\label{amplify}
\end{figure}


\subsection{Amplification}

The aforesaid discussion gives an idea of how the two modules work together to
generate the pool of $A_{2P}$ out of the total pool of BvgA. Through western blot technique
\cite{Prugnola1995} gave a qualitative picture of amplification of the BvgA protein due to 
temperature shift. Guided by the experimental observation of \cite{Prugnola1995} and
considering the amplification is due to the accumulation of large amount of $A_{2P}$ within the 
cell (see the preceding discussion and Figure~(\ref{phase})), we now look at the response of 
the system in terms of accumulation of $A_{2P}$ as a function of external stimulus. To this end 
we use the concept of amplification factor, $A_s$ proposed by \cite{Koshland1982}. In our 
notation $A_s$ can be defined as
\begin{equation}
\label{afactor}
A_s = \frac{\Delta A_{2P} / A_{2P}^i}{\Delta k_{p,s2} / k_{p,s2}^i}
        = \frac{\left ( A_{2P}^f -A_{2P}^i \right )/A_{2P}^i}{\left ( k_{p,s2}^f-k_{p,s2}^i \right )/k_{p,s2}^i} ,
\end{equation}

\noindent
where $i$ and $f$ refer to the initial and final value, respectively, for the input parameter
$k_{p,s2}$ and output $A_{2P}$. Using Eq.~(\ref{afactor}) we have calculated the amplification 
factor as a function of input stimulus. The resultant data are plotted in Figure~\ref{amplify}
which shows a gradual amplification of TF as external stimulus is increased. From
Figure~\ref{amplify} it is also evident that due to high stimulus the gain raises upto 
$\sim 42 \%$ compared to its value at low signal.


\subsection{Time dependent dynamics}

\begin{figure}[!t]
\begin{center}
\includegraphics[width=0.75\columnwidth,angle=-90]{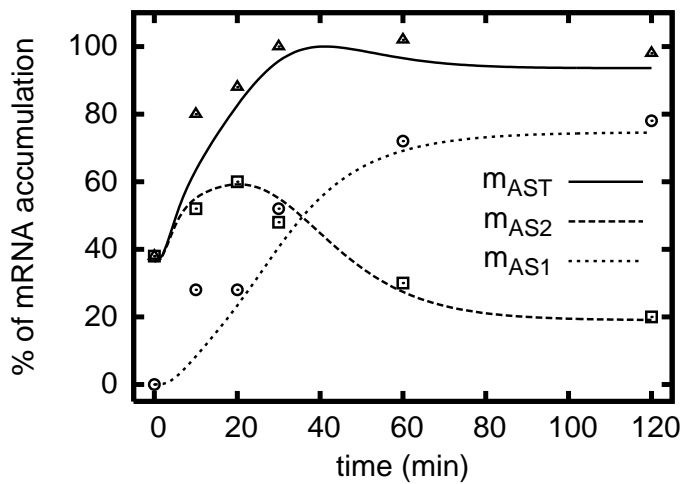}
\end{center}
\caption{Time evolution of transcripts $m_{AS1}$ and $m_{AS2}$ due to $P_{AS1}$
and $P_{AS2}$ activity, respectively. The open symbols are taken from 
\cite{Scarlato1991} and the solid lines are results of numerical integration.
}
\label{mrna}
\end{figure}

\begin{figure}[!b]
\begin{center}
\includegraphics[width=0.75\columnwidth,angle=-90]{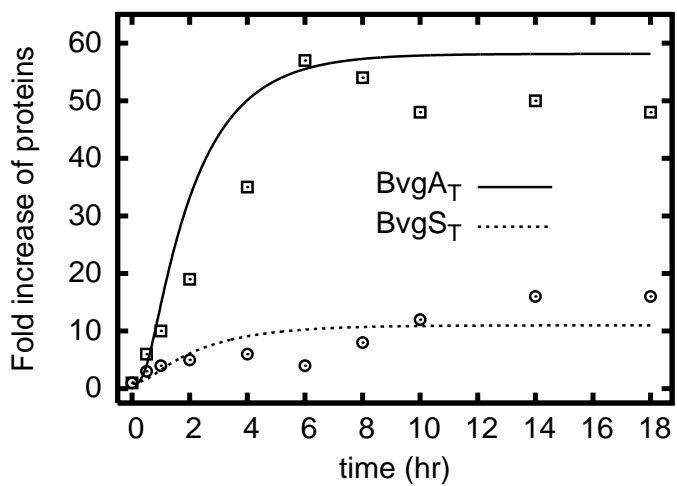}
\end{center}
\caption{Fold increase of total amount of sensor ($S_T$) and response
regulator ($A_T$) protein. The dynamics is shown for first 18 hours after induction.
The open symbols are taken from \cite{Scarlato1991} and the solid lines 
are results of numerical integration.
}
\label{protein}
\end{figure}

To study the time dependent behavior of different quantities of the model, the sets of 
nonlinear ODEs are solved by XPP (http://www.math.pitt.edu/$\sim$bard/xpp/xpp.html) 
using the parameter set given in Table~2. The parameters listed in Table~2 were guessed
to generate the temporal experimental profile given in Figures~(\ref{mrna}, \ref{protein},
\ref{dgene} and \ref{mut-1}a). The consensus set of parameters used for numerical integration 
of the nonlinear ODEs are obtained using Parameter Estimation Toolkit (PET) 
(http://mpf.biol.vt.edu/pet/).


\subsubsection{The {\it bvg} operon}

In Figure~\ref{mrna} we compare the numerical results with experimental data 
\citep{Scarlato1991},
for time evolution of transcripts $m_{AS1}$ and $m_{AS2}$ generated by the two promoters
P$_{AS1}$ and P$_{AS2}$, respectively. While plotting the data we have scaled each 
transcripts value
by the maximum of $m_{AST}$ $(=m_{AS1}+m_{AS2})$ to compare simulated
data with experimental results in a relative scale of 100. From \ref{mrna} it is evident that our model
captures the qualitative aspects of {\it in vivo} experimental results. To get an idea of how well
our model can mimic the real system we define the amplification factor, $f$
(= induced/basal) for mRNA expression  which for experiment and simulation are
$f_{exp}\approx 2.63$ and $f_{sim}=k_{tp,11}/k_{tp,20} \approx 2.15$, respectively, and
are in good agreement.


\subsubsection{The proteins}

As the {\it bvg} operon gets switched on at higher temperature (37 $^{\circ}$C) it starts producing
a large pool of response regulator BvgA which is about $\geqslant 50$ fold higher than that what is produced at the non-inducing condition \citep{Scarlato1990,Scarlato1991,Prugnola1995}.
Experimental results show low level synthesis of sensor (BvgS) and response regulator (BvgA)
proteins at low temperature (25 $^{\circ}$C), but 6 hours after induction their level becomes
56 and 4 fold higher, respectively. In the next 18 hours level of BvgA protein goes down to 40 fold
whereas level of BvgS protein finally increases to 17 fold \citep{Scarlato1991}. 
It is important to note that downfall of BvgA level after first 6 hours is a signature of 
arrival of {\it B. pertussis} colony at the stationary state where nutrition and other growth 
resources may become limited, that effectively lead to nonlinear degradation
\citep{Monod1949,Tan2009,Ghosh2011}. In the present model we have considered only
linear degradation kinetics as the relevant dynamics for the activation of the downstream genes
takes place within first 8 hours of induction (see discussion in the following). In other words,
within the first 6-8 hours the pool of BvgA becomes saturated enough to trigger the
signal transduction network. Keeping this in mind, the present model with linear degradation
kinetics can still take care of the relevant dynamics within the exponential growth phase and
at the transition from exponential growth phase to stationary phase.
In Figure~\ref{protein} we
show qualitative agreement of simulated result with that of experimental data for fold increase of
total BvgS ($S_T$) and total BvgA ($A_T$) protein.  While caculating $S_T$ and $A_T$ we have
used relations (Eq.~(14)-Eq.~(15)).


\begin{figure}[!t]
\begin{center}
\includegraphics[width=0.75\columnwidth,angle=-90]{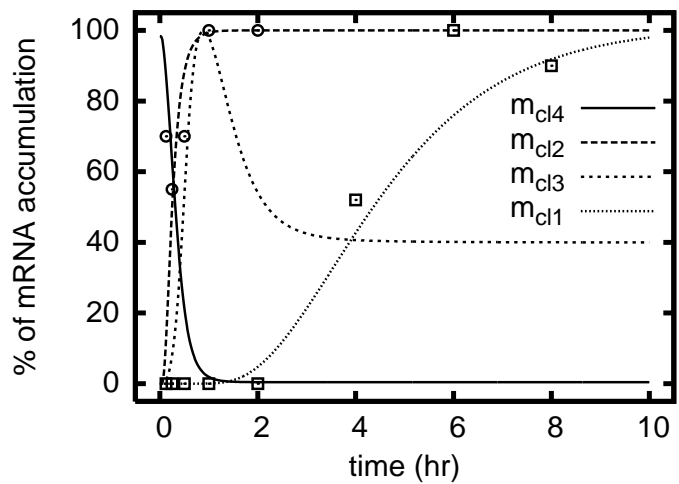}
\end{center}
\caption{Time evolution of four different classes of mRNA due to differential
gene regulation. The open symbols are taken from \cite{Scarlato1991}
and the solid lines are due to numerical integration.
}
\label{dgene}
\end{figure}


\subsubsection{Gene regulation of phenotypic phases}

After 2 hours of induction total BvgA level increases only by $\sim$18 fold,
but such low level of response regulator protein is enough to activate class 2 genes that
encode the proteins for adherence due to high affinity binding site upstream of TSP.
As a result, within 2 hours of induction the promoter for class 2 gene
gets activated and the corresponding mRNA ($m_{cl2}$) level reaches its maximum value
(see Figure~\ref{dgene}). After this 2 hours time window, level of BvgA rises and the accumulated amount
is enough to bind the low affinity binding site of class 1 genes and to activate them. Thus, in a
period of 2-6 hours of activation, class 1 genes gets fully {\it on} leading the corresponding
mRNA ($m_{cl1}$) level to its maximum value (see Figure~\ref{dgene}).

Together with class 1 and class 2 genes we have shown the time evolution of class 3 and
class 4 genes in Figure~\ref{dgene}. As mentioned in the introduction, although class 3 gene 
expression
has not been observed in {\it B. pertussis} under temperature induction, we predict that
proper modulation of external temperature may lead to expression of class 3 gene  in
{\it B. pertussis} as gradual tuning of external temperature slowly accumulates the
transcription factor in a graded response manner \citep{Prugnola1995}.
Similarly, expression of class 4 gene can be achieved by incorporating a plasmid containing
{\it frlAB} fused with {\it lacZ} from {\it B. bronchiseptica} into
{\it B. pertussis}. Similar kind of technique has been used to express {\it ptx} promoter
from {\it B. pertussis} in {\it B. bronchiseptica} \citep{Williams2007}.


\subsubsection{The mutants}

In the previous discussion we have provided an account of how well the developed
model can reproduce the qualitative features of signal transduction at the molecular
level. Next we check the validity of the model by looking into some novel
mutants reported in the literature \citep{Jones2005}. As mentioned earlier necessary
condition for the activation of signal transduction in {\it B. pertussis} is the phosphorylation
of the TF, BvgA. Thus, any form of hindrance/activation through mutation at the 
phosphorylation site of BvgA will get reflected in their ability to activate the signaling 
machinery. At this point it is important to mention that in an earlier \textit{in vitro} set up 
it has been observed that BvgA gets phosphate from GST tagged BvgS \citep{Uhl1994}.
Keeping this in mind two mutants, R152H and T194M along with the wild type
BvgA have been used to study an {\it in vitro} phosphorylation kinetics \citep{Jones2005} 
where in a mixture of 0.8 $\mu$M GST-$'$BvgS and 2.1 $\mu$M BvgA (wild type or R152H 
or T194M) 30 $\mu$M [$\gamma$-$^{32}$P]-ATP was added and the phosphotransfer
kinetics was monitored at 0.8, 2 and 5 minutes after addition. The relative amount of
radioactive phosphate incorporated into BvgA was then calculated as a function of time using 
phosphoimager (see Figures 5B and 5C of \cite{Jones2005}) in a relative scale of 100. 
The mutant R152H
has been reported to behave almost like wild type in their ability of getting phosphorylated
whereas T194M has been shown to be phosphorylated in a drastically low amount,
$\sim 20 \%$ of that of wild type BvgA.

\begin{figure}[!t]
\begin{center}
\includegraphics[width=0.75\columnwidth,angle=-90,bb=48 62 590 792]{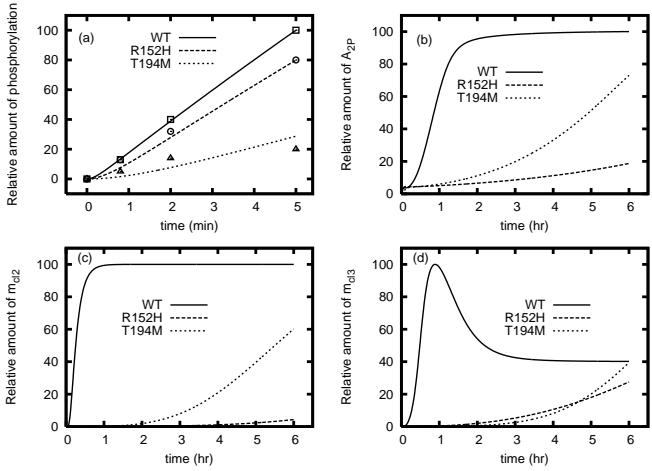}
\end{center}
\caption{Effect of mutation on phosphorylation of BvgA and its effect on the
activation of genes with high affinity binding site.
(a) Comparison of \textit{in vitro} phosphorylation data (symbols are from \cite{Jones2005})
and theoretical results (solid, dashed and dotted line) for WT, R152H and T194M.
(b) Relative amount of $A_{2P}$ due to WT and the two mutants R152H and T194M.
(c) and (d) Relative amount of transcripts generated from genes with high affinity
binding site of TF. 
In all the four panels all the data are scaled with respect to the maximum value,
set to 100.
}
\label{mut-1}
\end{figure}

In our model one can control the rate of phosphorylation of TF through the 
Michaelis constant $K_{Mt} (= (k_{t,b}+k_{t,a2})/k_{t,f})$ for the kinase
activity of the sensors. For the two mutants R152H and T194M we take $K_{Mt}$
to be 3.42 nM and 0.5 nM, respectively, compared to 9.96 nM for wild 
type strain. The {\it in vitro} phosphorylation assay results can be modeled in the 
present study using only Eqs.~(\ref{autop}-\ref{ptrans}), as Eq.~(\ref{autop}) mimics
phosphorylation of GST tagged BvgS using [$\gamma$-$^{32}$P]-ATP and
Eq.~(\ref{ptrans}) takes care of kinase activity of BvgS towards BvgA.
To reproduce the \textit{in vitro} kinetic data reported by \cite{Jones2005}
using our model we have used 0.8 $\mu$M $S_2$ and 2.1 $\mu$M $A_2$ (for wild type,
R152H and T194M) and varied the Michaelis constant $K_{Mt}$ which
mimics the role of radioactive ATP incorporation (from $S_2$ to $A_2$)
in the experimental setup. In Figure~\ref{mut-1}(a) we show the \textit{in vitro} 
phosphorylation assay results (symbols) along with data generated using 
Eqs.(\ref{autop}-\ref{ptrans}) (lines), which shows a good agreement between 
theory and experiment.

Nonspecific binding between high affinity binding site and BvgA can occur 
even when the later is unphosphorylated but its specific affinity for DNA 
increases in the phosphorylated form \citep{Boucher1997}. 
From this information one can expect phosphorylated BvgA with R152H 
substitution will have higher binding probability to the primary high 
affinity BvgA binding site, compared to T194M substitution. But surprisingly, 
almost reverse result was observed in electrophoretic mobility shift assays (EMSA)
\citep{Jones2005}. In the EMSA experiment an oligonucleotide (22SYM) with a 
perfect inverted heptameric repeat at the center was used as it has been shown to
represent a high affinity BvgA binding site earlier 
\citep{Boucher1995,Boucher1997,Boucher2001}. Resulting data shows that,
compared to wild type, BvgA with T194M substitution shows $\sim 40 \%$ binding 
ability compared to wild type. Whereas, BvgA with R152H substitution shows almost 
zero binding affinity towards the high affinity DNA binding site.
The {\it in vitro} phosphorylation assay and electrophoretic mobility shift assay
results show reverse effect for R152H and T194M when it comes to binding
probability to primary high affinity binding site. Thus, {\it in vitro} results suggests
that although replacement of arginine (R) by histidine (H) at 152 residue retains the positive 
charge, its interaction with negatively charged double helix backbone reduces
severely. On the other hand replacement of threonine (T) by methionine (M) at 194
position increases the hydrophobicity of the TF so that inspite of being poorly
phosphorylated their interaction with high affinity binding site increases.


The above mentioned results with reverse effects thus immediately raises the question -
how these two mutants will behave, if expressed {\it in vivo}? To answer this
we have predicted the temporal behavior of TF, class~2 mRNA and class~3
mRNA for R152H and T194M substitution and compared them with wild type
profiles (see Figure~\ref{mut-1}(b)-\ref{mut-1}(d)). While simulating the full network
(using Eqs.~(1-13)) we have decreased all the binding and unbinding 
($k_b$ and $k_u$) rates of the model, listed in Table~2, by two order of magnitude to 
get desired phenotypic behavior for the mutants R152H and T194M. 
Our {\it in silico} prediction suggests that opposing effect of phosphorylation and 
binding affinity causes delay in the activation of the signal transduction machinery 
as reported in \cite{Jones2005}.


\begin{figure}[!b]
\begin{center}
\includegraphics[width=0.75\columnwidth,angle=-90,bb=48 62 590 792]{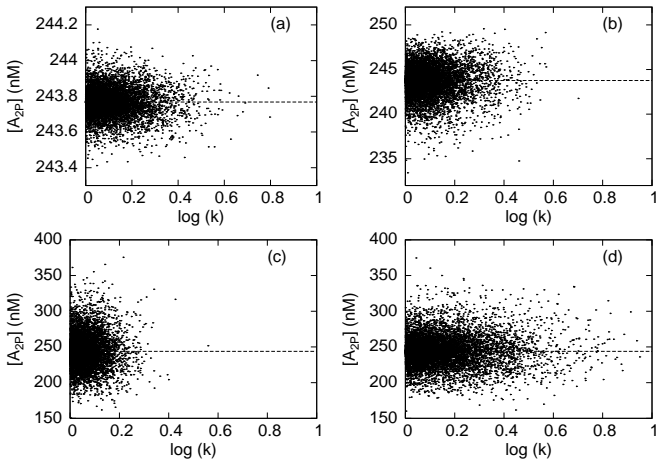}
\end{center}
\caption{Plot of steady state $[A_{2P}]$ as a function of total parameter variation
$\log (k)$. The horizontal dashed line is the steady state $A_{2P}$ value ($\sim 244$ nM)
for the reference system whereas each dot represents the same for modified set.
(a) Only binding constants are modified; (b) only synthesis and degradation rates
are modified; (c) only kinase and phosphatase rates are modified and (d)
all the rates are modified.
}
\label{sens}
\end{figure}


\begin{figure}
\begin{center}
\includegraphics[width=0.75\columnwidth,angle=-90]{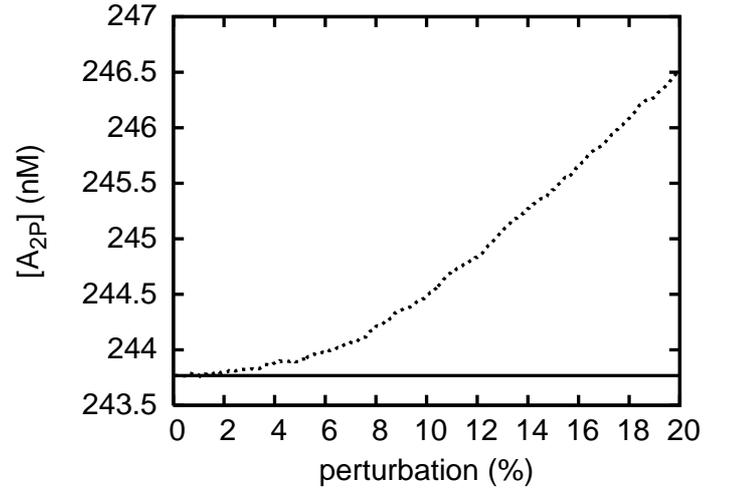}
\end{center}
\caption{Evolution of mean $[A_{2P}]$ level as function of perturbation (\%) (dotted line).
The horizontal solid line is the steady state $A_{2P}$ value ($\sim 244$ nM) for the
unperturbed system.
}
\label{pert}
\end{figure}


\subsubsection{Sensitivity Analysis}

To check the sensitivity of the parameter values on the network dynamics listed in 
Table~2 we have adopted the procedure described by \cite{Barkai1997}. In this 
procedure all or a subset of the rate constants were subjected to a random perturbation 
where the perturbation was drawn from a random gaussian distribution whose mean 
is the unperturbed value of each rate constant and variance is certain percentage 
(up to maximum of 10\%) of the rate constant. This leads to a set of unperturbed 
(reference) rate constants, $k_i^0$ (listed in Table~2) and an ensemble of perturbed 
(modified) set of 
rate constants, $k_i$. We then calculated the level of $A_{2P}$ at steady state using 
both the reference set and the modified set using the full network. Variation of the 
reference steady state $[A_{2P}]$ value for the modified set of rate constants can be 
characterized by total parameter variation $k$, defined as 
$\log (k) = \sum_{i=1}^N \left | \log (k_i/k_i^0) \right |$ \citep{Barkai1997}. The resultant 
simulation
results are showed in Figure~\ref{sens} where the dotted line is the steady state 
$[A_{2P}]$ for the reference state and each dot represents the same using the
perturbed set. We have tested the sensitivity of the parameters using four different sets.
In set 1 we have only perturbed the binding constants (Figure~\ref{sens}a); in set 2
the rate constants controlling the synthesis and the degradation of the systems components
have been modified (Figure~\ref{sens}b); in set 3 the rate constants for kinase and 
phosphatase activities have been changed (Figure~\ref{sens}c) and finally in set 4
all the rate constants given in Table~2 have been modified (Figure~\ref{sens}d). From
the sensitivity test it is evident that the rate constants responsible for kinase and
phosphatase activities are most sensitive to random perturbation (see the range of
ordinate in Figure~\ref{sens}c) compared to the other parameters of the model.

To check further whether the parameter set given in Table~2 can withstand larger perturbation
and gives physically realizable quantity we have perturbed all the parameters up to 20\%. 
Using the perturbation method mentioned before we have generated 1000 data set for each 
level of perturbation and calculated the mean $[A_{2P}]$ level to compare with the base value. 
In Figure~(\ref{pert}) we show the evolution of mean $[A_{2P}]$ level as a function of perturbation,
which shows that even with perturbation as large as 20\% the mean protein level (dotted 
line) lies within the 1\% of the unperturbed value (the solid horizontal line). Beyond 20\% 
perturbation the solution diverges by several orders of magnitude and becomes unrealistic.


\section{Conclusion}

The kinetic model developed in the present study takes care of signal transduction
and phenotypic gene regulation in {\it B. pertussis} under the influence of temperature
elevation. The proposed model includes all possible elementary kinetics of the biochemical
interactions between several system components. To understand the molecular switch
operative in the BvgAS TCS, a quasi steady state analysis has been performed which 
reveals temperature induced sharp molecular switch that responds to the external stimulus, 
a reminiscent of amplified sensitivity \citep{Goldbeter1981,Koshland1982}. Development of the 
sharp switch has been shown to be the consequence of positive feedback motif present 
within the {\it bvgAS} operon that becomes operative when the temperature 
of the surrounding is increased (Figure~\ref{amplify}). Outcome of the sharp switch gets 
reflected in the accumulation 
of the TF in large amount within few hours of induction (Figure~\ref{protein}), which then controls 
the expression of several downstream genes including the genes for adherence and toxin. 
All these features have been observed via numerical integration of the coupled nonlinear 
ODEs for large set of parameters. The resultant numerical results essentially capture the 
qualitative features of the network dynamics performed {\it in vivo}. On the basis of our 
developed model we then looked into the behavior of two novel mutants impaired in their ability
to phosphorylate the transcription factor and made testable predictions for temperature induced 
class 3 gene expression.

We hope that our {\it in silico} study will inspire more experiments in the coming days
to address subtle issues of the network that are yet to be explored. One of such issues is the
characterization of exact target and functioning of the antisense RNA, transcribed from
$P_{AS4}$ promoter of  {\it bvgAS} operon.  In this connection it is important to mention the
recent work by \cite{Hot2011}, where sRNA of \text{B. pertussis} have been identified
for the first time. Out of the pool of sRNA identified one (\textit{bprJ2}) has been mentioned 
to be controlled
by BvgAS TCS but exact target and mode of its functionality is yet to be discovered.
In addition, one may find it interesting to make quantitative measurement of different 
proteins generated due to activity of the four classes of downstream genes. Information 
from these new experimental findings will certainly help one to build a more detailed model 
in future.


\section*{Acknowledgements}

We express our sincerest gratitude to Indrani Bose, Sudip Chattopadhyay,
Gaurab Gangopadhyay and Sandip Kar for stimulating discussions. AB acknowledges 
Council of Scientific and Industrial Research (CSIR), Government of India for research 
fellowship (09/015(0375)/2009-EMR-I). SKB acknowledges support from Bose Institute 
through Institutional Programme VI - Development of Systems Biology.


\section{Appendix}

\subsection{Promoter kinetics of downstream genes}

The active and inactive forms of four different promoters of the downstream genes
are modeled as $P_{clj,a}$ and $P_{clj,i}$ ($j=1,2,3,4$), respectively. Considering
co-operativity in binding of TF ($A_{2P}$) to the high/low affinity binding
site of these promoters we model binding kinetics as follows,

\noindent For class 1 gene:
\begin{subequations}
\begin{eqnarray}
\label{cl1p}
P_{cl1,i} + A_{2P} \overset{k_{b,11}}{\underset{k_{u,11}}{\rightleftharpoons}}
P_{cl1,i1} , \\
P_{cl1,i1} + A_{2P} \overset{k_{b,12}}{\underset{k_{u,12}}{\rightleftharpoons}}
P_{cl1,i2} , \\
P_{cl1,i2} + A_{2P} \overset{k_{b,13}}{\underset{k_{u,13}}{\rightleftharpoons}}
P_{cl1,a} .
\end{eqnarray}
\end{subequations}

\noindent For class 2 gene:
\begin{subequations}
\begin{eqnarray}
\label{cl2p}
P_{cl2,i} + A_{2P} \overset{k_{b,21}}{\underset{k_{u,21}}{\rightleftharpoons}}
P_{cl2,i1} , \\
P_{cl2,i1} + A_{2P} \overset{k_{b,22}}{\underset{k_{u,22}}{\rightleftharpoons}}
P_{cl2,i2} , \\
P_{cl2,i2} + A_{2P} \overset{k_{b,23}}{\underset{k_{u,23}}{\rightleftharpoons}}
P_{cl2,a} .
\end{eqnarray}
\end{subequations}

\noindent For class 3 gene:
\begin{subequations}
\begin{eqnarray}
\label{cl3p}
P_{cl3,i} + A_{2P} \overset{k_{b,31}}{\underset{k_{u,31}}{\rightleftharpoons}}
P_{cl3,i1} , \\
P_{cl3,i1} + A_{2P} \overset{k_{b,32}}{\underset{k_{u,32}}{\rightleftharpoons}}
P_{cl3,a} , \\
P_{cl3,a} + A_{2P} \overset{k_{b,33}}{\underset{k_{u,33}}{\rightleftharpoons}}
P_{cl3,i2} , 
\end{eqnarray}
\end{subequations}

\noindent For class 4 gene:
\begin{equation}
\label{cl4p}
P_{cl4,a} + A_{2P} \overset{k_{b,41}}{\underset{k_{u,41}}{\rightleftharpoons}}
P_{cl4,i} ,
\end{equation}

\noindent
In the above set of equations $P_{clj,ik}$ $(k=1,2,\ldots)$ represents the inactive intermediate
states of the different classes of promoters.

\subsection{Promoter kinetics of bvg locus}

As mentioned in the main text we consider a single copy of the {\it bvg} gene, with
relations $[P_{AS2c}]+[P_{AS2a}]+[P_{AS2i}]=1$ and 
$[P_{AS1i}]+[P_{AS1a}]=1$ for the two
promoters controlling functionality of the {\it bvg} operon. Dynamical equations
for the promoter kinetics thus can be written as
\begin{eqnarray}
\label{eqa}
\frac{d[P_{AS2c}]}{dt} &=&   -k_{b2} [P_{AS2c}] [A_{2P}] + k_{u2} [P_{AS2a}] , \\
\label{eqa1}
\frac{d[P_{AS2a}]}{dt}  &=&  k_{b2} [P_{AS2c}] [A_{2P}] + k_{a2} [P_{AS2i}] \nonumber \\
&& - (k_{u2} + k_{i2}) [P_{AS2a}], \\
\label{eqa2}
\frac{d[P_{AS1a}]}{dt}  &=&  k_{b1} [P_{AS1i}] [A_{2P}] - k_{u1} [P_{AS1a}] ,
\end{eqnarray}

\noindent
which at the steady state give,
\begin{eqnarray}
\label{eqa3}
& [P_{AS1a}] = F_1 (A_{2P}), [P_{AS1i}] = 1 - F_1 (A_{2P}), \nonumber \\
& [P_{AS2c}] = F_2 (A_{2P}), [P_{AS2a}] = F_2 (A_{2P}) [A_{2P}]_2,
\end{eqnarray}

\noindent
where
\begin{eqnarray*}
F_1 (A_{2P}) = \frac{[A_{2P}]_1}{1+[A_{2P}]_1} ,
F_2 (A_{2P}) = \frac{1}{1+(1+k) [A_{2P}]_2},
\end{eqnarray*}

\noindent
with $[A_{2P}]_i = [A_{2P}]/K_i$ for $K_i=k_{ui}/k_{bi}$ ($i=1,2$) and $k=k_{i2}/k_{a2}$.

\subsection{mRNA kinetics}

Time evolution of the transcripts generated from the two promoters $P_{AS2}$ and 
$P_{AS1}$ are given by the following set of equations, respectively,
\begin{eqnarray}
\label{eqa4}
\frac{d[m_{AS2}]}{dt}  &=&  k_{tp,20} [P_{AS2c}] + k_{tp,21} [P_{AS2a}] \nonumber \\
&& - k_{d,m} [m_{AS2}] , \\
\label{eqa5}
\frac{d[m_{AS1}]}{dt}  &=&  k_{tp,11} [P_{AS1a}] - k_{d,m} [m_{AS1}] .
\end{eqnarray}

\noindent
Along with the steady state expression for the two promoters given in Eq.~(\ref{eqa3}),
one arrives at the following expressions for concentrations of the two transcripts
at steady state ($d[m_{AS2}]/dt = d[m_{AS1}]/dt = 0$)
\begin{eqnarray}
\label{eqa6}
&& [m_{AS2}] = \frac{k_{tp,20}}{k_{d,m}} F_2 (A_{2P})
+ \frac{k_{tp,21}}{k_{d,m}} F_2 (A_{2P}) [A_{2P}]_2, \nonumber \\
&&[m_{AS1}]  = \frac{k_{tp,11}}{k_{d,m}} F_1 (A_{2P}) .
\end{eqnarray}

\subsection{Phosphotransfer kinetics}

From the kinetics of phosphate donation from sensor to
response regulator (kinase), and phosphate withdrawal from response regulator
by sensor (phosphatase), we have two dynamical equations for the two intermediates
$S_{2P}\cdot A_2$ and $S_2\cdot A_{2P}$
\begin{eqnarray}
\label{eqa7}
\frac{d[S_{2P}\cdot A_2]}{dt} &=& k_{t,f} [S_{2P}][A_2] \nonumber \\
&& - (k_{t,b}+k_{t,a2}) [S_{2P}\cdot A_2] , \\
\label{eqa8}
\frac{d[S_2\cdot A_{2P}]}{dt} &=& k_{p,f} [S_2][A_{2P}] \nonumber \\
&&- (k_{p,b}+k_{p,a2}) [S_2\cdot A_{2P}] .
\end{eqnarray}

\noindent
While writing the above two equations we have neglected degradation of the intermediates
as they show transient dynamics, i.e., time evolution of both the Michalies complexes takes
place on a faster time scale compared to the other system components.
Now by imposing quasi-steady state conditions 
$d[S_{2P}\cdot A_2]/dt=0$ and $d[S_2\cdot A_{2P}]/dt=0$ on
Eqs.~(\ref{eqa7}-\ref{eqa8}) we have
\begin{equation}
\label{eqa9}
[S_{2P}\cdot A_2] = \frac{[S_{2P}][A_2]}{K_{Mt}} ,
[S_2\cdot A_{2P}] = \frac{[S_2][A_{2P}]}{K_{Mp}},
\end{equation}

\noindent
where $K_{Mt} = (k_{t,b}+k_{t,a2})/k_{t,f}$ and $K_{Mp} = (k_{p,b}+k_{p,a2})/k_{p,f}$ are
the Michaelis constants for the kinase and the phosphatase activity of the sensors, respectively.

\subsection{Protein kinetics}

The dynamical equations representing kinetics of the different forms of
sensor and response regulator proteins can be represented by the following sets of
ordinary differential equations

\begin{eqnarray}
\label{eqa10}
\frac{d[S_2]}{dt}  & =  & \tilde{\alpha}_1 F_2 (A_{2P})
+ \tilde{\alpha}_2 F_2 (A_{2P}) [A_{2P}]_2
+ \tilde{\alpha}_3 F_1 (A_{2P}) \nonumber \\
&& - k_{p,s2}(s) [S_2] + k_{dp,s2} [S_{2P}]  
+ \frac{k_{t,a2}}{K_{Mt}} [S_{2P}][A_2]  \nonumber \\
&&- k_{d,p} [S_2] , \\
\label{eqa11}
\frac{d[S_{2P}]}{dt}  & =  & k_{p,s2}(s) [S_2] - k_{dp,s2} [S_{2P}] -
\frac{k_{t,a2}}{K_{Mt}} [S_{2P}] [A_2] \nonumber \\
&& - k_{d,p} [S_{2P}] , \\
\label{eqa12}
\frac{d[A_2]}{dt}  & =  & \tilde{\beta}_1 F_2 (A_{2P})
+ \tilde{\beta}_2 F_2 (A_{2P}) [A_{2P}]_2
+  \tilde{\beta}_3 F_1 (A_{2P}) \nonumber \\
&& -\frac{k_{t,a2}}{K_{Mt}} [S_{2P}][A_2]
+ \frac{k_{p,a2}}{K_{Mp}} [S_2][A_{2P}] \nonumber \\
&& - k_{d,p} [A_2] , \\
\label{eqa13}
\frac{d[A_{2P}]}{dt}  & =  & \frac{k_{t,a2}}{K_{Mt}} [S_{2P}] [A_2]
- \frac{k_{p,a2}}{K_{Mp}} [S_2] [A_{2P}] \nonumber \\
&& - k_{d,p} [A_{2P}] ,
\end{eqnarray}

\noindent
where
\begin{eqnarray*}
\tilde{\alpha}_1 = k_{ss,2} \frac{k_{tp,20}}{k_{d,m}} ,
\tilde{\alpha}_2 = k_{ss,2} \frac{k_{tp,21}}{k_{d,m}} ,
\tilde{\alpha}_3 = k_{ss,1} \frac{k_{tp,11}}{k_{d,m}} , \\
\tilde{\beta}_1 = k_{sa,2} \frac{k_{tp,20}}{k_{d,m}} ,
\tilde{\beta}_2 = k_{sa,2} \frac{k_{tp,21}}{k_{d,m}} ,
\tilde{\beta}_3 = k_{sa,1} \frac{k_{tp,11}}{k_{d,m}} .
\end{eqnarray*}

\noindent
While writing Eqs.~(\ref{eqa10}-\ref{eqa13}) we have used the steady state expressions for
$[m_{AS1}], [m_{AS2}], [S_{2P}\cdot A_2]$ and $[S_2\cdot A_{2P}]$ given by Eq.~(\ref{eqa6}) 
and Eq.~(\ref{eqa9}).

\subsection{Linear stability analysis}

Using the relations $[S_2]+[S_{2P}] \approx [S_T]$,
$[A_2]+[A_{2P}] \approx [A_T]$ and Eqs.~(\ref{eqa10}-\ref{eqa13})
one can write
\begin{eqnarray}
\label{eqa14}
\frac{d[S_T]}{dt} & = & f(A_{2P}) - k_{d,p} [S_T]  = 0, \\
\label{eqa15}
\frac{d[A_T]}{dt}  & = & g(A_{2P}) - k_{d,p} [A_T]  = 0,
\end{eqnarray}

\noindent
where
\begin{eqnarray}
\label{eqa16}
f(A_{2P}) & = & \tilde{\alpha}_1 F_2 (A_{2P})
+ \tilde{\alpha}_2 F_2 (A_{2P}) [A_{2P}]_2 \nonumber \\
&& + \tilde{\alpha}_3 F_1 (A_{2P}) , \\
\label{eqa17}
g(A_{2P}) & =  & \tilde{\beta}_1 F_2 (A_{2P})
+ \tilde{\beta}_2 F_2 (A_{2P}) [A_{2P}]_2 \nonumber \\
&& +  \tilde{\beta}_3 F_1 (A_{2P}) .
\end{eqnarray}

\noindent
At 25 $^\circ$C when the switch is off, $A_{2P}=0$ and hence 
$f(A_{2P})= \tilde{\alpha}_1 + \tilde{\alpha}_3,$ and 
$g(A_{2P})= \tilde{\beta}_1 + \tilde{\beta}_3,$. Thus we have
at steady state the fixed point
\begin{eqnarray*}
[S_T]_{ss} = \alpha_1,
[A_T]_{ss} = \beta_1
\end{eqnarray*}

\noindent
where $\alpha_i = \tilde{\alpha}_i/k_{d,p}$ and $\beta_i = \tilde{\beta}_i/k_{d,p}$
for $i=1-3$.
On the other hand, at 37 $^\circ$C for $A_{2P}^{ss}$, the steady state value of $A_{2P}$,
we have
\begin{eqnarray*}
[S_T]_{ss}^* = \frac{1}{k_{d,p}} f(A_{2P}^{ss}),
[A_T]_{ss}^* = \frac{1}{k_{d,p}} g(A_{2P}^{ss}).
\end{eqnarray*}

\noindent
To understand the nature of the fixed point $([S_T]_{ss}^*,[A_T]_{ss}^*)$ 
we construct the stability matrix evaluated at steady state,
\begin{eqnarray*}
\left (
\begin{array}{cc}
-k_{d,p} & 0 \\
0 & -k_{d,p}
\end{array}
\right ) ,
\end{eqnarray*}

\noindent
for which trace, $\tau = -2 k_{d,p} (<0)$ and determinant, $\Delta = k_{d,p}^2 (>0)$,
a characteristics of stable fixed point.




\onecolumn


\begin{center}
\begin{longtable}{lll}
\caption{\label{lssymb} List of symbols (with initial values) used in the model} \\
\hline
Symbol & Initial Value & Description \\
\hline
$P_{AS1a}$  &  0 nM &  Active state of promoter $P_{AS1}$ \\
$P_{AS1i}$   &  0.98 nM &  Inactive state of promoter $P_{AS1}$ \\
$P_{AS2c}$  &  0.98 nM &  Constitutive state of promoter $P_{AS2}$ \\
$P_{AS2a}$   &   0 nM &  Active state of promoter $P_{AS2}$ \\
$P_{AS2i}$   &   0 nM &  Inactive state of promoter $P_{AS2}$ \\
$m_{AS1}$ & 0 nM & Transcripts generated from $P_{AS1a}$ \\
$m_{AS2}$ & 1.11 nM & Transcripts generated from $P_{AS2a}$ \\
$S_2$ & 10.74 nM & Dimer of sensor BvgS \\
$A_2$ & 11.23 nM & Dimer of response regulator BvgA \\
$S_{2P}$ & 0 nM & Phosphorylated dimer of sensor BvgS \\
$A_{2P}$ & 0 nM & Phosphorylated dimer of response \\
                   & & regulator BvgA (transcription factor) \\
$S_{2P}\cdot A_2$ & 0 nM & Michaelis complex formed by $S_{2P}$ and $A_2$ \\
$S_2\cdot A_{2P}$ & 0 nM & Michaelis complex formed by $S_2$ and $A_{2P}$ \\
$P_{cl1,a}$ &  0 nM & Active promoter of class 1 gene \\
$P_{cl1,i}$ &   0.98 nM & Inactive promoter of class 1 gene \\
$P_{cl2,a}$ &  0 nM & Active promoter of class 2 gene \\
$P_{cl2,i}$ &   0.98 nM & Inactive promoter of class 2 gene \\
$P_{cl3,a}$ &  0 nM & Active promoter of class 3 gene \\
$P_{cl3,i}$ &   0.98 nM & Inactive promoter of class 3 gene \\
$P_{cl4,a}$ &  0.98 nM & Active promoter of class 4 gene \\
$P_{cl4,i}$ &   0 nM & Inactive promoter of class 4 gene \\
$m_{cl1}$ & 0 nM & Transcripts generated from $P_{cl1,a}$ \\
$m_{cl2}$ & 0 nM & Transcripts generated from $P_{cl2,a}$ \\
$m_{cl3}$ & 0 nM & Transcripts generated from $P_{cl3,a}$ \\
$m_{cl4}$ & 2.93 nM & Transcripts generated from $P_{cl4,a}$ \\
\hline
\end{longtable}
\end{center}

\newpage


\begin{center}
\begin{longtable}{lll}
\caption{\label{prmwt} List of kinetic parameters (with values) used in the model} \\
\hline
Parameter & Value & Description \\
\hline
$k_{b1}$  &  $1.024 \times 10^{-4}$ nM$^{-1}$ s$^{-1}$  & Association rate of $A_{2P}$ 
and $P_{AS1i}$ \\
$k_{u1}$  &  $1.167 \times 10^{-3}$ s$^{-1}$  & Dissociation rate of $A_{2P}$ from $P_{AS1a}$ \\
$k_{b2}$  &  $1.36 \times 10^{-3}$ nM$^{-1}$ s$^{-1}$  & Association rate of $A_{2P}$ and 
$P_{AS2c}$ 
\\
$k_{u2}$  &  $2.5 \times 10^{-2}$      s$^{-1}$  & Dissociation rate of $A_{2P}$ from $P_{AS2a}$ \\
$k_{i2}$  &  $1.667 \times 10^{-3}$   s$^{-1}$  & Inactivation of $P_{2a}$ to $P_{AS2i}$ \\
$k_{a2}$  & $2.0 \times 10^{-4}$   s$^{-1}$  & Activation of $P_{2a}$ from $P_{AS2i}$ \\
$k_{tp,20}$ &  $1.9 \times 10^{-3}$ s$^{-1}$  & Transcription rate from $P_{AS2c}$ promoter \\
$k_{tp,21}$ &  $9.386 \times 10^{-3}$ nM$^{-1}$ s$^{-1}$  & Transcription rate from 
$P_{AS2a}$ promoter \\
$k_{tp,11}$ &  $4.083 \times 10^{-3}$ s$^{-1}$  & Transcription rate from 
$P_{AS1a}$ promoter \\
$k_{d,m}$ &  $1.667 \times 10^{-3}$ s$^{-1}$  & Degradation rate of mRNA \\
$k_{ss,1}$ &  $6.667 \times 10^{-3}$ s$^{-1}$  & Synthesis rate of $S_2$ from $m_{AS1}$ \\
$k_{ss,2}$ &  $1.667 \times 10^{-3}$ s$^{-1}$  & Synthesis rate of $S_2$ from $m_{AS2}$ \\
$k_{sa,1}$ &  $4.167 \times 10^{-2}$ s$^{-1}$  & Synthesis rate of $A_2$ from $m_{AS1}$ \\
$k_{sa,2}$ &  $1.667 \times 10^{-3}$ s$^{-1}$  & Synthesis rate of $A_2$ from $m_{AS2}$ \\
$k_{p,s2}(s)$ &  $8.333 \times 10^{-3}$ s$^{-1}$  & Phosphorylation rate of $S_2$ at 37 $^{\circ}$C\\
$k_{dp,s2}$ &  $3.333 \times 10^{-3}$ s$^{-1}$  & Dephosphorylation rate of $S_{2P}$ \\
$k_{t,f}$ &  $8.532 \times 10^{-3}$ nM$^{-1}$ s$^{-1}$  & Association rate of $S_{2P}$ and $A_2$ \\
$k_{t,b}$ &  $1.667 \times 10^{-3}$ s$^{-1}$  & Dissociation rate of $S_{2P}\cdot A_2$ \\
$k_{t,a2}$ &  $8.333 \times 10^{-2}$ s$^{-1}$  & Phosphate transfer rate from $S_{2P}$ to $A_2$ \\
$k_{p,f}$ &  $3.413 \times 10^{-5}$ nM$^{-1}$ s$^{-1}$  & Association rate of $S_2$ and $A_{2P}$ \\
$k_{p,b}$ &  $1.333 \times 10^{-3}$ s$^{-1}$  & Dissociation rate of $S_2\cdot A_{2P}$ \\
$k_{p,a2}$ &  $5.0 \times 10^{-2}$ s$^{-1}$  & Phosphate removal rate from $A_{2P}$ by $S_2$ \\
$k_{d,p}$ &  $1.667 \times 10^{-4}$ s$^{-1}$  & Degradation rate of protein \\
$k_{b,11}$  &  $6.826 \times 10^{-7}$ nM$^{-1}$ s$^{-1}$  & Association rate of $A_{2P}$ and $P_{cl1,i}$ \\
$k_{u,11}$  &  $1.667 \times 10^{-6}$ s$^{-1}$  & Dissociation rate of $A_{2P}$ from $P_{cl1,i1}$ \\
$k_{b,12}$  & $1.024 \times 10^{-6}$ nM$^{-1}$ s$^{-1}$  & Association rate of $A_{2P}$ and $P_{cl1,i1}$ \\
$k_{u,12}$  &  $1.667 \times 10^{-6}$ s$^{-1}$  & Dissociation rate of $A_{2P}$ from $P_{cl1,i2}$ \\
$k_{b,13}$  &  $1.36 \times 10^{-6}$ nM$^{-1}$ s$^{-1}$  & Association rate of $A_{2P}$ and $P_{cl1,i2}$ \\
$k_{u,13}$  &  $1.667 \times 10^{-6}$ s$^{-1}$  & Dissociation rate of $A_{2P}$ from $P_{cl1,a}$ \\
$k_{b,21}$  &  $5.119 \times 10^{-4}$ nM$^{-1}$ s$^{-1}$  & Association rate of $A_{2P}$ and $P_{cl2,i}$ \\
$k_{u,21}$  &  $1.667 \times 10^{-4}$ s$^{-1}$  & Dissociation rate of $A_{2P}$ from $P_{cl2,i1}$ \\
$k_{b,22}$  &  $1.36 \times 10^{-3}$ nM$^{-1}$ s$^{-1}$  & Association rate of $A_{2P}$ and $P_{cl2,i1}$ \\
$k_{u,22}$  &  $1.667 \times 10^{-4}$ s$^{-1}$  & Dissociation rate of $A_{2P}$ from $P_{cl2,i2}$ \\
$k_{b,23}$  & $1.706 \times 10^{-3}$ nM$^{-1}$ s$^{-1}$  & Association rate of $A_{2P}$ and $P_{cl2,i2}$ \\
$k_{u,23}$  &  $1.667 \times 10^{-4}$ s$^{-1}$  & Dissociation rate of $A_{2P}$ from $P_{cl2,a}$ \\
$k_{b,31}$  & $8.533 \times 10^{-5}$ nM$^{-1}$ s$^{-1}$  & Association rate of $A_{2P}$ and $P_{cl3,i}$ \\
$k_{u,31}$  &  $1.667 \times 10^{-4}$ s$^{-1}$  & Dissociation rate of $A_{2P}$ from $P_{cl3,i1}$ \\
$k_{b,32}$  &  $1.365 \times 10^{-4}$ nM$^{-1}$ s$^{-1}$  & Association rate of $A_{2P}$ and $P_{cl3,i1}$ \\
$k_{u,32}$  &  $1.667 \times 10^{-4}$ s$^{-1}$  & Dissociation rate of $A_{2P}$ from $P_{cl3,a}$ \\
$k_{b,33}$  &  $1.706 \times 10^{-6}$ nM$^{-1}$ s$^{-1}$  & Association rate of $A_{2P}$ and $P_{cl3,a}$ \\
$k_{u,33}$  &  $2.0 \times 10^{-4}$ s$^{-1}$  & Dissociation rate of $A_{2P}$ from $P_{cl3,i2}$ \\
%
$k_{b,41}$  &  $1.706 \times 10^{-4}$ nM$^{-1}$ s$^{-1}$  & Association rate of $A_{2P}$ and $P_{cl4,a}$ \\
$k_{u,41}$  &  $1.667 \times 10^{-4}$ s$^{-1}$  & Dissociation rate of $A_{2P}$ from $P_{cl4,i}$ \\
$k_{tp,cl1}$ &  $5.167 \times 10^{-3}$ s$^{-1}$  & Transcription rate from $P_{cl1,a}$ promoter \\
$k_{tp,cl2}$ &  $5.083 \times 10^{-3}$ s$^{-1}$  & Transcription rate from $P_{cl2,a}$ promoter \\
$k_{tp,cl3}$ &  $6.16 \times 10^{-3}$ s$^{-1}$  & Transcription rate from $P_{cl3,a}$ promoter \\
$k_{tp,cl4}$ &  $5.0 \times 10^{-3}$ s$^{-1}$  & Transcription rate from $P_{cl4,a}$ promoter \\
\hline
\end{longtable}
\end{center}

\end{document}